\documentclass[preprint,aps,floatfix,showpacs]{revtex4}
\usepackage{graphicx}

\newcommand{\bea}{\begin{eqnarray}}
\newcommand{\eea}{\end{eqnarray}}
\newcommand\beq{\begin{equation}}
\newcommand\eeq{\end{equation}}
\newcommand\beqa{\begin{eqnarray}}
\newcommand\eeqa{\end{eqnarray}}

\def\gtap{ \rlap{$>$}\lower5pt\hbox{$\sim$}}

\def\tr{\mathrm{Tr}}
\def\mui{\mu_I}

\def\OMIT#1{}

\usepackage{amssymb}

\begin{document}
\vspace{0.5cm}
\title{\phantom{x}
\vspace{0.5cm}  Skyrmions in the presence of isospin chemical potential}

\author{
J. A. Ponciano$^{a,b}$ \thanks{e-mail: ponciano@fisica.unlp.edu.ar} and
N. N. Scoccola$^{c,d,e}$ \thanks{e-mail: scoccola@tandar.cnea.gov.ar}}

\affiliation{
$^a$ CEFIMAS, Av. Santa Fe 1145, (1059) Buenos Aires, Argentina.\\
$^b$ Universidad Nacional de La Plata, C.C. 67, (1900), La Plata, Argentina.\\
$^c$ Physics Department, CNEA,
     (1429) Buenos Aires, Argentina.\\
$^d$ CONICET, Rivadavia 1917, (1033) Buenos Aires, Argentina.\\
$^e$ Universidad Favaloro, Sol{\'\i}s 453, (1078) Buenos Aires,
Argentina.}

\vspace*{1.cm}

\date{\today}

\vspace*{1.cm}

\begin{abstract}
We analyze the existence of localized finite energy topological
excitations on top of the perturbative pion vacuum
within the Skyrme model at finite isospin chemical potential and
finite pion mass. We show that there is a critical isospin
chemical potential $\mu_I^c$ above which such solutions cease to
exist. We find that $\mu_I^c$ is closely related to the value of
the pion mass. In particular for vanishing pion mass we obtain
$\mu_I^c=0$ in contradiction with some results recently reported
in the literature. We also find that below $\mui^c$ the skyrmion mass
and baryon radius show, at least for the case of the hedgehog ansatz,
only a mild dependence on the isospin chemical potential.
\end{abstract}

\maketitle


Hadronic systems with vanishing baryon chemical potential $\mu_B$
and finite isospin chemical potential $\mu_I$ are unstable with
respect to weak decays. However, if we are interested in the
dynamics of the strong interaction alone, we can disregard the
relative slow electroweak effects and consider them as stable.
Moreover, although there are not yet precise lattice QCD
calculations at finite baryon density due to the Fermion sign
problem, it is in principle possible to perform lattice
simulations at finite isospin density\cite{Alford:1998sd}. These
remarks have led several groups to study the behavior of strongly
interacting matter at $\mu_B=0$ and finite $\mu_I$. Effective
lagrangian analysis showed that there is a phase transition from
normal phase to pion superfluidity at a critical isospin chemical
potential which turns out to be equal to the pion mass in the
vacuum\cite{Son:2000xc}. This has been confirmed by lattice QCD
calculations\cite{Kogut:2002tm}, random matrix method
analysis\cite{Klein:2003fy}, etc. Studies at finite temperature
have been also performed\cite{Loewe:2002tw}. Given these results
it is of considerable interest to investigate on the behavior of
baryon properties at finite isospin chemical potential. One model
which is well suited to perform these studies is the Skyrme model.
In the Skyrme model\cite{Skyrme:1961vq} and its generalizations,
baryons arise as topological excitations of a non-linear chiral
Lagrangian written in terms of meson fields. These  type of models
have been quite successful in describing the properties of octet
and decouplet baryons (see e.g., Refs.\cite{Zahed:1986qz}). In a series of
recent articles\cite{Loewe:2004ic} the skyrmion properties in the
presence of the isospin chemical potential have been analyzed. It
has been found that, in the case of vanishing pion mass, there is
a critical chemical potential $\mu_I^c \approx 223\ MeV$ above
which stable soliton solutions cease to exist. Moreover, according
to Refs. \cite{Loewe:2004ic} the skyrmion mass vanishes at $\mu =
\mu_I^c$. In this letter we re-examine the issue of the skyrmion
stability for finite isospin chemical potential considering the
possibility of having a finite pion mass. We show that although
there is indeed a critical isospin chemical potential $\mu_I^c$,
its value is closely related to the value of the pion mass, and
for vanishing pion mass one has $\mu_I^c=0$. Moreover we find
that, at least within the spherically symmetric hedgehog ansatz,
the skyrmion mass and baryon radius remain rather stable up to the
critical isospin chemical potential.


We start by considering the lagrangian of the $SU(2)$ Skyrme model
with quartic term stabilization and finite pion mass. It is given
by
\begin{eqnarray}
{\cal L} =
- \frac{f_\pi^2}{4} \tr \left\{ L_\alpha L^\alpha \right\}
+ \frac{1}{32e^2} \tr \left\{ [L_\alpha,L_\beta]^2 \right\}
+ \frac{m_\pi^2 \ f_\pi^2}{4} \tr\left\{ U + U^\dagger - 2 \right\}.
\label{skyr}
\end{eqnarray}
Here, $f_\pi$ is the pion decay constant whose empirical value is
$f_\pi^{emp} = 93\ MeV$, $e$ is the so-called Skyrme parameter and
$m_\pi$ is the pion mass which we will take at its empirical value
$m_\pi^{emp} = 139\ MeV$. In our numerical calculations below we
will use the standard set of values $f_\pi = 54 \ MeV$ and $e =
4.84$ which leads to the empirical value of the nucleon and
$\Delta$ masses within the rigid rotor approximation\cite{AN84}.
It should be stressed, however, that our main conclusions are
expected to be independent of this particular choice of
parameters. In Eq.(\ref{skyr}), as usual, $U$ represents the
$SU(2)$ chiral field and the Maurier-Cartan operator $L_\alpha$ is
defined by $L_\alpha = U^\dagger
\partial_\alpha U$.

The isospin chemical potential $\mui$ is introduced by performing
the replacement
\begin{equation}
\partial_\alpha U \longrightarrow \partial_\alpha U -
i \ \frac{\mui}{2} \ [\tau_3 , U] \ g_{\alpha 0},
\end{equation}
where $g_{\alpha\beta}$ is the metric tensor in Minkowski space
and $\tau_3$ is the third Pauli matrix. This leads to a modified
lagrangian which reads
\begin{eqnarray}
{\cal L} (\mu_I) = {\cal L} + \frac{\mui^2 f_\pi^2}{16} \tr\left\{
\omega^2 \right\} - \frac{\mui^2}{64 e^2} \tr\left\{
[\omega,L_\alpha]^2 \right\} \! + \! \frac{i \mui f_\pi^2}{4} \tr\left\{
\omega L_0 \right\} - \frac{i \mui }{8 e^2} \tr\left\{ \omega
L_\alpha [ L_0 , L_\alpha ] \right\}
\end{eqnarray}
where $\omega = U^\dagger \tau_3 U - \tau_3$.

In what follows we will be interested in static soliton configurations.
The corresponding soliton mass reads
\begin{eqnarray}
M(\mui) &=& - \int d^3x \left[ \frac{f_\pi^2}{4} \tr \left\{ L_i
L_i \right\} + \frac{1}{32e^2} \tr \left\{ [L_i,L_j]^2 \right\}
+ \frac{m_\pi^2 \ f_\pi^2}{4} \tr\left\{ U + U^\dagger - 2 \right\} \right. \nonumber \\
& & \left. \qquad \qquad + \frac{\mui^2 f_\pi^2}{16} \tr\left\{
\omega^2 \right\} + \frac{\mui^2}{64 e^2} \tr\left\{
[\omega,L_i]^2 \right\} \right].
\end{eqnarray}

It is not hard to see that the terms proportional to $\mui$ are
not invariant under isospin rotations. Namely, the isospin
chemical potential introduces a preferred direction in isospin
space which is expected to lead to an axially deformed soliton
configuration. For the time being we will assume that such
deformations are small for the range of values of $\mui$
considered here.
Therefore we introduce the usual spherically symmetric
hedgehog ansatz for the baryon number $B=1$ configuration
\beq
U_H
= \exp\left[ i \vec \tau \cdot \hat r \ F(r) \right].
\eeq
In this case we obtain
\begin{eqnarray}
M_H(\mui) & = & \frac{f_\pi^2}{2} \int d^3x \left[ F'^2 + 2 \ \frac{s^2}{r^2}
\left( 1 + \frac{F'^2}{e^2 f_\pi^2} \right) + \frac{1}{e^2
f_\pi^2} \frac{s^4}{r^4}
+ 2 \ m_\pi^2 \left( 1 - c \right) \right. \nonumber \\
& & \qquad \qquad \left. - \frac23 \ \mui^2 \ s^2 \left( 1 + \frac{F'^2}{e^2 f_\pi^2} \right)
- \frac23 \ \mui^2 \ \frac{s^4}{r^2} \right],
\end{eqnarray}
where $s = \sin F$ and $c = \cos F$. The minimization of $M_H(\mui)$
leads to the following Euler-Lagrange equation for the soliton profile $F$
\begin{eqnarray}
& & F'' \left[ 1 + \frac{2\ s^2}{e^2 f_\pi^2 r^2}  \left( 1 - \frac{\mui^2\ r^2}{3} \right) \right]
+ 2 \ \frac{F'}{r} \left( 1 - \frac{ 2 \ \mui^2\ s^2}{3} \right)  - \frac{2\ s\ c}{r^2}
\left( 1 - \frac{F'^2}{e^2 f_\pi^2} \right) \left( 1 - \frac{\mui^2 \ r^2}{3} \right)
\nonumber \\
& &
\qquad  - \frac{2 \ s^3 \ c}{e^2 f_\pi^2 r^4} \left(1 - \frac23 \ \mui^2 \ r^2 \right) - m_\pi^2\ s
= 0.
\label{ele}
\end{eqnarray}
As usual this differential equation is supplemented by the boundary conditions
corresponding to $B=1$ topological
excitations on top of the perturbative pion vacuum, $F(0) = \pi$ and $F(\infty)=0$.
The associated baryon radius is given by
\begin{eqnarray}
r_B(\mui) = \left( -\frac2\pi \int_0^\infty dr\ r^2 \ s^2 \ F' \right)^{1/2}.
\end{eqnarray}

Before presenting the numerical soliton solutions we will analyze the behavior of the
profile $F(r)$ for large distances. Given the boundary conditions we can linearize
Eq.(\ref{ele}) in that limit. We obtain
\begin{eqnarray}
& & F'' + \frac{ 2 F'}{r} - \left( m_\pi^2 - \frac23 \ \mui^2 + \frac{2}{r^2}
\right) F = 0.
\label{ele2}
\end{eqnarray}
This equation implies that localized finite energy topological
excitations on top of the perturbative pion vacuum exist only for
$\mui \leq \mui^c$, where $\mui^c = \sqrt{3/2}\ m_\pi$. For $\mui
> \mui^c$ the solutions have an oscillatory behavior. This is
clearly displayed in Fig.1 where we plot the numerical solutions
of Eq.(\ref{ele}) for some chosen values of $\mui$. In fact, the
situation is quite similar to the one found long time ago in the
study of the stability of the skyrmion under spin-isospin
rotations\cite{Braaten:1984qe}. Note that for $\mui = \mui^c$ the
solution displays a $1/r^2$ large distance behavior which is
typical of the localized pion massless case.

In the region where localized solutions exist we can study the behavior of the soliton
mass and baryon radius as a function of the isospin chemical potential. This is shown in
Fig.2. We observe that as $\mui$ increases, the soliton mass decreases while the radius
increases.
However, these effects are not too large (less than 5 \% at $\mui = \mui^c$),
and $M_H(\mui)$ never vanishes in such region.

As mentioned at the beginning of this letter, general arguments indicate
that in the meson sector there is a phase transition from the normal
phase (perturbative pion vacuum) to a pion condensed phase at
$\mui^c =  m_\pi$. Comparing with the result obtained above we note that
in the soliton sector it appears an extra factor $\sqrt{3/2}$ in the
corresponding critical value. A similar factor has been found in the
study of the stability of the hedgehog skyrmion under spin-isospin
rotation\cite{Battye:2005}. As in that case, it reflects the fact
that in the hedgehog approximation possible pion excitations are
assumed to be spherical while, as already mentioned, the presence
of the isospin chemical potential is expected to induce axially
symmetric deformations. In order to account for this fact
we introduce a general axial ansatz which, in cylindrical polar
coordinates $(\rho,~\phi,z)$, is given by\cite{Battye:2005,Krusch:2004},
\begin{equation}
U_{ax}=\psi_3+i\tau_3 \psi_2 +i \psi_1 (\tau_1 \cos \phi + \tau_2 \sin \phi ),
\end{equation}
Here, $\psi_a$ are the components of a unit vector $\vec \psi (\rho,z)$ that is independent of the angular variable $\phi$.
The boundary conditions for finite energy solutions are that $\vec \psi \rightarrow (0,0,1)$ as $\rho^2 + z^2 \rightarrow \infty$
and that on the symmetry axis $\rho =0$ the equations $\psi_1=0$ and $\partial_\rho \psi_2=\partial_\rho \psi_3=0$ must be satisfied.
Using this ansatz, the mass for the static soliton configuration reads
\begin{eqnarray}
 M_{ax}(\mui)&=&\pi f_\pi^2\int  \left[ \partial_i \vec \psi \cdot \partial_i \vec \psi
 \left(1+\frac{1}{e^2f_\pi^2}\frac{\psi_1^2}{\rho^2}\right) +
 \frac{1}{e^2f_\pi^2}|\partial_\rho \vec \psi \times \partial_z \vec \psi|^2 + \frac{\psi_1^2}{\rho^2} \right. \nonumber \\
& & \left. + 2 m^2_\pi(\psi_3-1) -\mui^2 \psi_1^2\left( 1+ \frac{1}{e^2f_\pi^2} (|\vec \psi \times \partial_z \vec \psi |^2 +
|\vec \psi \times \partial_\rho \vec \psi|^2)\right) \right] \rho~d\rho ~dz,
\end{eqnarray}
where $i=\rho,z$. Minimizing this expression we obtain a set of
coupled equations for the two independent functions that we take
to be $\psi_1(\rho,z)$ and $\psi_2(\rho,z)$. Unfortunately the
resolution of these equations implies a rather time-consuming
numerical task. From the results obtained in the case of spin-isospin
rotations\cite{Battye:2005}, soliton properties are not expected
to be too much affected by the axial deformations provided $\mui \leq \mui^c$.
Thus we postpone this numerical analysis for a future work. It is
important, however, to consider the linearized form of the
equations for $\psi_1$ and $\psi_2$, which are valid at large
distances. They read
\begin{eqnarray}
\partial_i \partial_i \psi_1 - \left( m_\pi^2 - \mui^2 + \frac{1}{\rho^2}\right)
\psi_1 & = & 0 \ , \nonumber \\
\partial_i \partial_i \psi_2 -  m_\pi^2 \ \psi_2 & = & 0 \ .
\end{eqnarray}
Thus, once axially symmetric configurations are considered, we
have indeed $\mui^c = m_\pi$. Namely, for $\mui \leq \mui^c$
the behavior of the deformed skyrmion mass and radius as a function
of the isospin chemical potential is expected to be very similar to
that shown in Fig.2, the only important difference being that the curve
will end at $\mui/m_\pi = 1$ instead of $\mui/m_\pi = \sqrt{3/2}$.

In conclusion, we have re-examined the behavior of the skyrmion
properties as a function of the isospin chemical potential $\mui$.
We have found that there is, indeed, a critical value of $\mui$
above which localized finite energy topological
excitations on top of the perturbative pion vacuum cease to exist. Such
critical value is closely related to the value of the pion mass.
For the spherically symmetric hedgehog ansatz we find $\mui^c =
\sqrt{3/2} \ m_\pi$, while when axially symmetric deformations are
allowed we obtain $\mui^c = \ m_\pi$ as expected from general
arguments in the meson sector. These results disagree with
previous works\cite{Loewe:2004ic} which report $\mu_I^c \approx
223 MeV$ in the chiral limit ($m_\pi=0$). Moreover, for $\mui \leq \mui^c$
we find only a mild dependence of the soliton mass and radius on $\mui$. In
particular, in contrast to the results in Ref.\cite{Loewe:2004ic} our values of
the soliton mass do not vanish for any value of $\mui \leq
\mui^c$. It should be noted that these analyses of the
soliton mass and radius dependence on $\mui$ have been performed using the
spherically symmetric hedgehog ansatz. Although for that range of values
of $\mui$ the axially symmetric deformations are expected to have only a minor effect
on this dependence, the numerical resolution of the corresponding
equations is required in order to confirm this expectation. This
will also allow us to explore the region $\mui > \mui^c$. We
hope to report on these issues in forthcoming publications.

\vspace*{.5cm}

{\it This work has been supported in part by CONICET and ANPCyT
(Argentina), under grants PIP 6084 and PICT04-03-25374,
respectively.}

\vspace*{2cm}

\begin{figure}[h]
\includegraphics[width=0.7\textwidth]{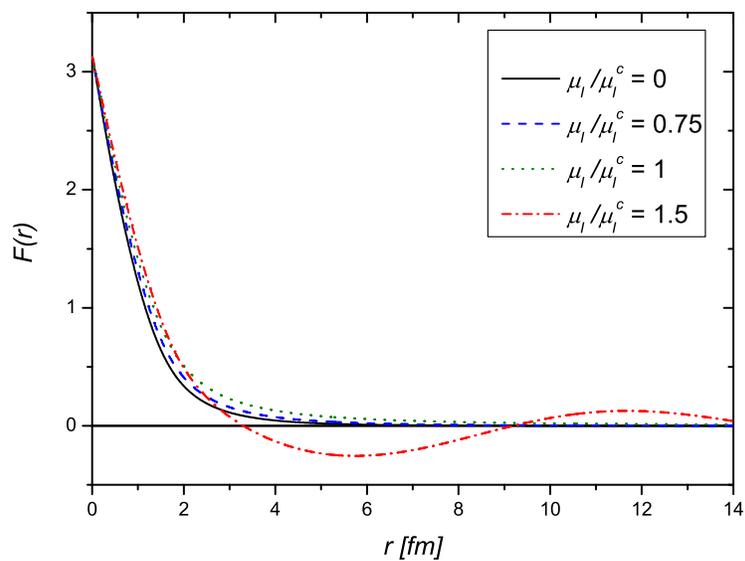}
\caption{(Color online) Hedgehog profiles $F(r)$ for various
values of the chemical potential $\mui$. Note that localized
solutions only exist for $\mui/\mui^c \leq 1$.}
\end{figure}

\begin{figure}[h]
\includegraphics[width=0.7\textwidth]{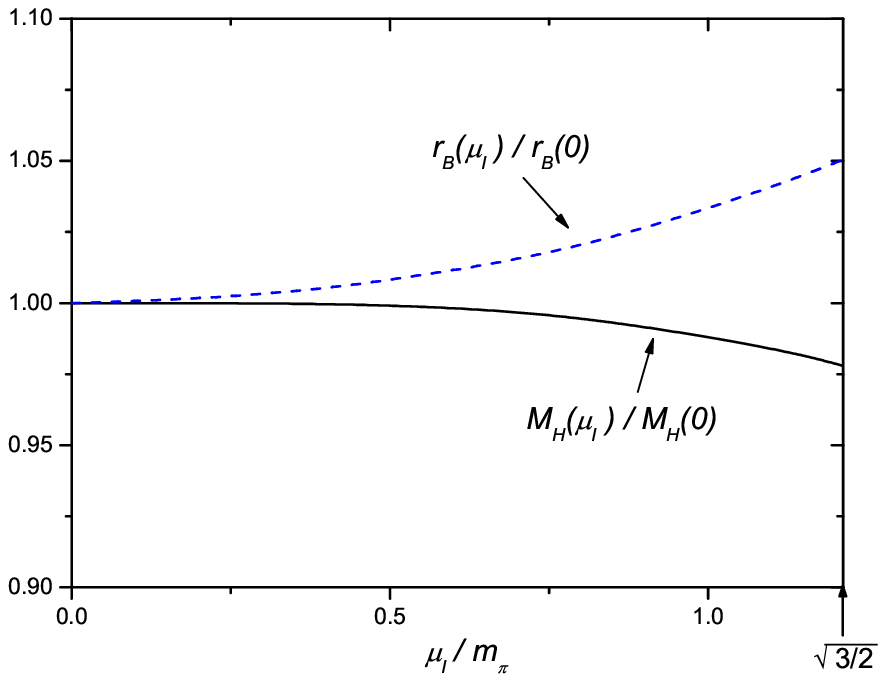}
\caption{(Color online) Soliton mass (full line) and baryon radius
(dashed line) as a function of the isospin chemical potential
$\mui$ for the spherically symmetric hedgehog ansatz and taken
with respect to their corresponding values at $\mui=0$. For our
parameter choice such values are $M_H(0) = 864 \ MeV$ and $r_B(0)
= 0.68 \ fm$\cite{AN84}.}
\end{figure}


\begin{thebibliography}{99}

\bibitem{Alford:1998sd}
M.~G.~Alford, A.~Kapustin and F.~Wilczek,
Phys.\ Rev.\  D {\bf 59} (1999) 054502.

\bibitem{Son:2000xc}
D.~T.~Son and M.~A.~Stephanov,
Phys.\ Rev.\ Lett.\  {\bf 86} (2001) 592;
J.~B.~Kogut and D.~Toublan,
Phys.\ Rev.\  D {\bf 64} (2001) 034007;
K.~Splittorff, D.~Toublan and J.~J.~M.~Verbaarschot,
Nucl.\ Phys.\  B {\bf 639} (2002) 524.

\bibitem{Kogut:2002tm}
J.~B.~Kogut and D.~K.~Sinclair,
Phys.\ Rev.\  D {\bf 66} (2002) 014508;
J.~B.~Kogut and D.~K.~Sinclair,
Phys.\ Rev.\  D {\bf 66} (2002) 034505.

\bibitem{Klein:2003fy}
B.~Klein, D.~Toublan and J.~J.~M.~Verbaarschot,
Phys.\ Rev.\  D {\bf 68} (2003) 014009.


\bibitem{Loewe:2002tw}
M.~Loewe and C.~Villavicencio,
Phys.\ Rev.\  D {\bf 67} (2003) 074034;
M.~Loewe and C.~Villavicencio,
Phys.\ Rev.\  D {\bf 70} (2004) 074005;
  M.~Loewe and C.~Villavicencio,
  Phys.\ Rev.\  D {\bf 71} (2005) 094001.





\bibitem{Skyrme:1961vq}
T.~H.~R.~Skyrme,
Proc.\ Roy.\ Soc.\ Lond.\  A {\bf 260} (1961) 127;
Nucl.\ Phys.\  {\bf 31} (1962) 556.

\bibitem{Zahed:1986qz}
I.~Zahed and G.~E.~Brown,
Phys.\ Rept.\  {\bf 142} (1986) 1;
H.~Weigel,
Int.\ J.\ Mod.\ Phys.\  A {\bf 11} (1996) 2419.

\bibitem{Loewe:2004ic}
M.~Loewe, S.~Mendizabal and J.~C.~Rojas,
Phys.\ Lett.\  B {\bf 609} (2005) 437;
Phys.\ Lett.\  B {\bf 632} (2006) 512;
Phys.\ Lett.\  B {\bf 638} (2006) 464.

\bibitem{AN84}
G.S. Adkins and C.R. Nappi,
Nucl. Phys. B {\bf 233} (1984) 109.


\bibitem{Braaten:1984qe}
E.~Braaten and J.~P.~Ralston,
Phys.\ Rev.\  D {\bf 31} (1985) 598;
R.~Rajaraman, H.~M.~Sommermann, J.~Wambach and H.~W.~Wyld,
Phys.\ Rev.\  D {\bf 33} (1986) 287.

\bibitem{Battye:2005}
R.~A.~Battye, ~S.~Krusch,~P.~M.~Sutcliffe,
Phys.\ Lett.\ B {\bf 626} (2005) 120.

\bibitem{Krusch:2004}
S.~Krusch, ~P.~M.~Sutcliffe,
J.\ Phys.\ A {\bf 37} (2004) 9037.

\end{thebibliography}
\end{document}